\documentclass[9pt,technote]{IEEEtran}

\usepackage{amssymb}
\usepackage{amsfonts}
\usepackage{amsthm}
\usepackage[utf8]{inputenc}
\usepackage{caption}
\usepackage{subcaption}
\usepackage[margin=1in]{geometry}
\usepackage{graphicx}
\usepackage{float}
\usepackage{rotating}
\usepackage{amsmath}
\usepackage{enumitem}
\usepackage{xcolor}
\usepackage{multicol}
\usepackage{hyperref}
\usepackage[ruled]{algorithm2e}
\hypersetup{
    colorlinks=true,allcolors=blue
}
\usepackage[sorting=none,style=science,citestyle=numeric]{biblatex}
\renewcommand{\vec}[1]{\mathbf{#1}}

\usepackage{graphics}

\title{Node Failure Localisation Problem for Load Balancing Dynamic Networks}
\date{}
\author{Ashley Barnes, Matthew Hole}
\begin{document}
\maketitle

\section*{Abstract}
Network tomography has been used as an approach to the Node Failure Localisation problem, whereby misbehaving subsets of nodes in a network are to be determined. Typically approaches in the literature assume a statically routed network, permitting linear algebraic arguments. In this work, a load balancing, dynamically routed network is studied, necessitating a stochastic representation of network dynamics. A network model was developed, permitting a novel application of Markov Chain Monte Carlo (MCMC) inference to the Node Failure Localisation (NFL) problem, and the assessment of monitor placement choices. Two nuanced monitor placement algorithms, including one designed for the NFL problem by Ma et al. 2014 were tested, with the published algorithm performing significantly better.  


\section{Introduction}
Network tomography, much like the X-ray Tomography used in medical imaging, is the inferring of internal structures, behaviours and properties of a network by analysing traffic passing between nodes in a network. Where X-ray analyses the behaviour of positrons passing through tissue to image the internal structure, network tomography can describe the inference of traffic flow, node status and structure of a network. Lawrence et al. \cite{lawrenceNetworkTomographyReview2006} discusses two broad themes in the study of network tomography: passive tomography where data is aggregated before inference is performed, and active tomography, in which one probes a network as it's in operation. 

The first studies in network tomography were largely passive, with Vardi et al. \cite{y.vardiNetworkTomographyEstimating2007} proposing a novel approach to the sender-receiver traffic problem. A Bayesian approach was then applied by \cite{tebaldiBayesianInferenceNetwork1998}, demonstrating that inference techniques can help to overcome the under-determined nature of many network tomography problems. Such monitor placement algorithms spread monitors throughout the network in a way that maximises their effectiveness, and minimises the number of monitors needed. The typical metric for measuring effectiveness for the node failure localisation is the maximum number of failed nodes that are able to be identified with a given number of monitors. 

With some exceptions, like the passive study of unicast networks, \cite{tsangPassiveNetworkTomography2001}, passive tomography broadly relies on data being collected on nodes or edges along the entire network. Naturally, this introduces significant challenges to implementation in real large scale networks. Conversely, an active tomography approach supposes that only a subset of nodes are under control of a network administrator, with probing traffic sent between such `monitor' nodes providing the dataset to be analysed. Whilst in theory more feasible for real world application, active tomography introduces a research question of its own: that of optimal monitor placement, and the focus of this research.

\subsection{Monitor Placement and the Failed Node Localisation Problem}
Given a network of known structure, the question is of where one should place monitor nodes so as to maximise the effectiveness of network tomography analysis. Ma et al. \cite{maNodeFailureLocalization2014} proposed a novel monitor placement algorithm for the failed node localisation problem. This placement algorithm approach has featured heavily in the literature in recent years, with further works extending the original algorithm to account for changes to network tomography \cite{heRobustEfficientMonitor2017}, proposing an algorithm designed for inferring city road traffic \cite{zhangNetworkTomographyApproach2018}, and relaxing assumptions about network reliability and taking a topological approach to algorithm design \cite{renRobustNetworkTomography2016}. 

Naturally, the algorithm being designed depends on the particular network tomography research question being asked. For instance, our research focusses on a modified version of the node failure localisation problem. In the original problem as studied by Ma et al., we suppose that on some network $G$ with nodes $N$ there are a subset of failed nodes $F \subset N$ which do not allow for the passage of traffic. 

\begin{figure}[H]
    \centering
    \includegraphics[width = 0.5\textwidth]{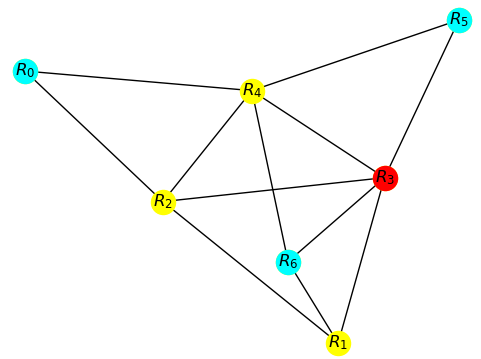}
    \caption{Example configuration for the Node Failure Localisation problem. Packets sent between blue monitor nodes can probe the network state, and data from such packets are then to be used to infer which of the nodes is not behaving correctly, in this case the red coloured router 3.}
    \label{figure:nodefailureexample}
\end{figure}

In Figure \ref{figure:nodefailureexample}, there are three yellow normal nodes, three blue monitor nodes and a red \textit{failed} node, which does not pass on packets that it receives. Ma et al. \cite{maNodeFailureLocalization2014} made the fixed routing assumption, fixing set of paths which can be probed by our monitor nodes. These paths are contained in the routing matrix $R$. In this example, we then have the set of paths $\overrightarrow{R_0R_5},\overrightarrow{R_0R_6},\overrightarrow{R_6R_5}$ and their reverse directions. It was then supposed that a series of probing packets was sent along each of these $m$ paths through a network with $n$ nodes. Define the binary vectors $\vec{p} \in \mathbb{2}^m$ and $\vec{q} \in \mathbb{2}^n$ such that

\begin{align}
    &p_i = 
    \begin{cases}
        1, \text{ if path }i\text{ fails}\\
        0, \text{ if path }i\text{ succeeds}
    \end{cases}\\
    &q_i = 
    \begin{cases}
        1, \text{ if node }i\text{ has failed}\\
        0, \text{ if node }i\text{ is functioning}
    \end{cases}
\end{align}

If one or more of the nodes traversed along $p_i$ had failed, then the probing packet would not be received, and the entire packet would fail. However, the fact that a path has failed only tells us that there is at least one failed node on this path. Similarly to the traffic intensity problem, we now have an inverse problem which can be written as an under-determined linear system. 

\begin{equation}
    \vec{p} = R \vec{q}
    \label{equation:failednode}
\end{equation}

Given a network configuration and a set of monitor nodes, Ma described the maximum identifiability of the network as the largest number of failed nodes which could be uniquely identified from probe path data without degeneracies in the solution. This maximum identifiability then provides a metric with which to measure the effectiveness of a monitor placement algorithm. 

Despite the variety of problems and approaches in the literature, one similarity in most network tomography studies is the linear algebraic nature of the problems. Network traffic is typically represented by a linear equation of the form of equation 

\begin{align}
    R\vec{x} = \vec{b}
    \label{equation:linearalgebra}
\end{align}

where $\vec{x}$ is a vector of unknown quantities, for instance edge/node states or edge traffic, a routing matrix $R$ dictating the paths to be taken between any pair of nodes, and a vector of measured quantities $\vec{b}$, corresponding to path state or edge traffic. This approach has proven to be versatile in its applicability to diverse research questions, but does still have limitations. Most prominently, as individual elements traversing the network are not resolved, phenomena involving the interaction of packets is difficult to represent. Queueing behaviour is one such unresolved feature using this method, meaning that dynamic load balancing and bottlenecks are not typically included in these network models. 

In this study of the monitor placement problem, the linear algebraic model was replaced by a stochastic network model, enabling the representation of more complex networking behaviour.

\subsection{Dynamic Load Balancing}
\label{section:loadbalancing}
When choosing the path to take between two nodes in a network, one might initially select the shortest path. However, in a network with sufficiently high traffic, we expect bottlenecks to occur stochastically, reducing the efficiency of the initial choice of routing paths. Intuitively, one might then choose to alter routing paths dynamically in order to choose the fastest route given current traffic conditions. This behaviour, familiar to drivers who use modern internet map services, is known as dynamic load balancing. Naturally, this feature is prominent in studies on transport networks \cite{rahmanBidirectionalTrafficManagement2015}, and is becoming increasingly relevant in computer networks with the rise in popularity of software defined networks (SDNs). Although not yet widely used on the internet, SDNs are a developing technology, and an attractive option for smaller scale local networks due the level of control afforded to the network administrators, and are likely to play a larger role in the future \cite{kreutzSoftwareDefinedNetworkingComprehensive2014, jiangLoadBalancingMulticasting2011}. Although more traditional routing protocols like Open Shortest Path First (OSPF) can't typically implement load balancing, many SDN protocols have been specifically designed to do so \cite{handigolPlugnServeLoadBalancingWeb2009,jiangLoadBalancingMulticasting2011,kreutzSoftwareDefinedNetworkingComprehensive2014}. The relevance of dynamic load balancing to both traffic and computer networks then motivates this study's stochastic approach to the monitor placement problem.

In incorporating a notion of time dependence and individual objects traversing the network stochastically, the node failure localisation problem was also altered. Instead of a binary `entirely failed' or `entirely working' distinction between nodes, the problematic nodes in our study took longer on average to process items in the queue. Inferring which nodes were problematic could not then rely on the binary failure states of individual paths as in \cite{heRobustEfficientMonitor2017} \cite{maNodeFailureLocalization2014}, but in comparing the distributions of traversal time and path lengths between sets of monitors. As the linear algebraic techniques for determining the sets of problematic routers were no longer applicable, Markov Chain Monte Carlo (MCMC) was employed to search the solution space.

\subsection{MCMC for Node Failure Localisation}
\label{section:MCMC}
MCMC was first used for network tomography by Tebaldi and West \autocite{tebaldiBayesianInferenceNetwork1998} in investigating the traffic origin-destination problem. The problem consists of large set of possible solutions, whose likelihoods can be directly computed and compared. In the stochastic node failure localisation problem, observed delay and path length distributions can be compared with those expected from candidate sets of problematic nodes. Suppose $\mathcal{D}^{\overrightarrow{ab}}_1$ and $\mathcal{D}^{\overrightarrow{ab}}_2$ are normalised packet delay distributions between monitors $a$ and $b$ for two sets of problematic nodes, denoted $1$ and $2$. To compare the two distributions, we take the mean squared error $\epsilon$, defined as

\begin{equation}
    \epsilon(A,B) = \frac{1}{n} \sum^{N}_{n = 1} (A_n - B_n)^2
\end{equation}.

To incorporate all other distributions from paths between other monitors, we define 

\begin{equation}
    \begin{aligned}
    \nu(\mathcal{D}_1,\mathcal{D}_2) := max ( &\epsilon(\mathcal{D}^{\overrightarrow{ab}}_1 , \mathcal{D}^{\overrightarrow{ab}}_2) , \\
    &\epsilon(\mathcal{D}^{\overrightarrow{ac}}_1 , \mathcal{D}^{\overrightarrow{ac}}_2) \dots \\
    &\epsilon(\mathcal{D}^{\overrightarrow{bc}}_1 , \mathcal{D}^{\overrightarrow{bc}}_2) ) 
    \end{aligned}
\end{equation}

The function $\nu$ then finds the largest error between two sets of distributions corresponding to different problematic node configurations. 

Suppose that in order to investigate the locations of problematic nodes on a network of known structure, the traffic between monitors has been measured, giving a set of observed distributions $\mathcal{D_O}$. Through a model of the network, the expected distributions associated with candidate sets of problematic nodes $\mathcal{D_C}$ can be calculated, and compared with the observed distribution. Given the large size of the solution space for even a modestly sized 20 node network, calculating $\nu(\mathcal{D_O},\mathcal{D_C})$ for each possible candidate configuration quickly becomes infeasible. However, by using $\nu$ in the Metropolis Hastings step of the MCMC Algorithm \ref{algorithm:MCMC}, the solution space can be explored more efficiently.

\begin{algorithm}
    \SetAlgoLined
    \textbf{Inputs:} 
    Observed distribution $\mathcal{D_O}$,Initial guess $\mathcal{C}_{old}$
    Calculate $\mathcal{D}_{\mathcal{C}_{old}}$
    \For{Number of Samples}{
        Choose an adjacent candidate $\mathcal{C}_{new}$
        Calculate $\mathcal{D}_{\mathcal{C}_{new}}$
        \underline{Metropolis-Hastings Step}\\
        \If{random variable $\chi_n \in [0,1] < max(1,\frac{\nu(\mathcal{D}_{\mathcal{T}},\mathcal{D}_{\mathcal{C}_{old}}}{\nu(\mathcal{D}_{\mathcal{T}},\mathcal{D}_{\mathcal{C}_{new}})})$}{
        \textit{$\mathcal{C}_{new}$ is accepted.}
        Save $\mathcal{C}_{new}$ and $\mathcal{D}_{\mathcal{C}_{new}}$ elsewhere
        Update $\mathcal{C}_{old} = \mathcal{C}_{new}$ and $\mathcal{D}_{\mathcal{C}_{old}} = \mathcal{D}_{\mathcal{C}_{new}}$
     }
     }
     \textbf{Output: }Set of points in the solution space explored
     \caption{MCMC for Node Failure Localisation}
     \label{algorithm:MCMC}
\end{algorithm}

The notion of adjacency in the solution space is defined as two configurations that differ only in one node's status. For example, the configuration $\{1,3\}$ is adjacent to both $\{1\}$ and $\{1,3,4\}$ but not with $\{2,3,4\}$.


\section{Model}
The need for time dependence and queueing behaviour necessitated a deviation from the typical linear algebraic treatment of network tomography. Instead, we developed a stochastic, agent based model. Each node in the network - whether a router or road intersection - behaves as an autonomous agent, managing its own queue and making routing decisions based on the queue states of other nodes in the network. For simplicity, we refer to the items traversing the network as `packets', although for the reader interested in transport networks this could easily be substituted with `vehicles'. Background packets originate stochastically at nodes on the network, with randomised target destinations, whilst probing packets are sent periodically along the `probing paths' between monitor nodes. The distributions of delay times and number of nodes traversed along each probing path is then the data used for tomographical analysis. 

Although this approach seeks to represent network traffic in a less abstracted manner than a linear algebraic formulation, it still makes a number of simplifying assumptions. 

\begin{enumerate}
    \item \textit{All nodes which are not nefarious behave identically.} 
    \item \textit{Background traffic across the network has a constant average intensity} 
    \item \textit{Inter-arrival time of new background packets is exponentially distributed } 
    \item \textit{All nodes are equally likely to send packets, and be chosen to receive packets}. 
    \item \textit{All routing protocols are the same for non-nefarious routers}
    \item \textit{The service time for every packet at the front of the queue in a router is one timestep}
    \item \textit{All traffic follows the same load balancing routing protocol}
    \item \textit{After the queue at a node fills, any subsequent packets sent to the queue are dropped}
    \item \textit{Each edge takes one timestep to traverse}

\end{enumerate}

Many of these assumptions pertain to homogeneity of nodes and behaviours across the network, which drastically simplify the analysis of results produced. If one were to use this model to study a specific network, these assumptions could be relaxed to better reflect the non-homogeneity of a real network, but as this study focusses on arbitrary, random networks, such specificity was deemed unnecessarily complex. The exponential inter-arrival time of packets reflects the work of Garsva et al. \cite{garsvaPacketInterarrivalTime2014}, who proposed this as a reasonably good fit for traffic across computer networks. The assumption of a finite queue length is a realistic one, in that for a network with sufficiently high network traffic we do not expect to see infinite queue lengths, the notion of packets being `dropped' is perfectly reasonable for computer networks, but naturally not what one would expect in a traffic network. 

Each node is an autonomous agent that keeps track of variables, namely:

\begin{itemize}
    \item A finite queue of packets waiting to be sent $\mathcal{Q}$
    \item A set of neighbouring routers $\mathcal{N}$
    \item A routing table $R$, dictating which immediate neighbour to forward outgoing packets to
    \item An address $A$.
\end{itemize}

In line with load balancing, the routing table $R$ is updated dynamically to best reflect the optimal routing paths for the current state of the network. We use the Extended Dijkstra's algorithm as suggested by Jiang et al. \cite{jiangLoadBalancingMulticasting2011} to be an ideal candidate for load balancing. The edge weights $W_{a,b}$ between every pair of nodes $a$ and $b$ as used by Dijkstra's algorithm are dependent on the queue lengths of neighbouring nodes such that 

\begin{align}
    W_{a,b} = 1 + \frac{\mathcal{Q}_a + \mathcal{Q}_b}{2}
    \label{equation:edgeweights}
\end{align}

With this implementation of load balancing, stochastically generated bottlenecks are avoided by subsequent packets, which are always routed in so as to avoid busy nodes. 

Packets are objects passed along the network, and they too, have associated variables:

\begin{itemize}[itemsep=1\itemsep,parsep=1\parsep,partopsep=1\partopsep,topsep=1\topsep]
    \setlength\itemsep{0.2cm}
    \item Address of sender $S$
    \item Address of intended recipient $R$
    \item Number of timesteps in transit $\tau$
    \item Number of nodes traversed $n$
\end{itemize}

After the simulation has finished running, distributions of $n$ and $\tau$ along each probing path are stored and used as the `observed' data, of the analogue of $\vec{b}$ in equation \ref{equation:linearalgebra}.


Each autonomous node undertakes a sequence of operations in each timestep as outlined in algorithm \ref{algorithm:routers}.

\begin{algorithm}[H]
    \textbf{Input}: Set of incoming packets $\mathcal{I}$. Routing table $R$\\
    \If{$\mathcal{Q} \neq \emptyset$}{
        \For{$p \in \mathcal{Q}$}{
            $p_{\tau} = p_{\tau} + 1$ \tcp*[l]{Increase the timesteps of waiting packets}
        }
        $p^{out} \leftarrow \mathcal{Q}_1$ \tcp*[l]{First packet in queue to be sent} 
        $\mathcal{Q} = \mathcal{Q} \backslash \{\mathcal{Q}_1\}$ \tcp*[l]{Remove first entry from queue}
        \If{$p^{out}_R = A$}{
            $p^{out}$ received by this node}
        \Else{
        Send $p^{out}$ to $R(A,p^{out}(A_r))$ 
        }
    }
    Append $\mathcal{I}$ to $\mathcal{Q}$ 
    \caption{Step Node Forwards in Time}
    \label{algorithm:routers}
\end{algorithm}

Algorithm \ref{algorithm:routers} allows the routers to behave autonomously, and pass packets along the network via their interactions with one another. Here we have laid the foundations for the agent based model. We now need a way to introduce traffic into the network.

\subsection{Stochastic Network Traffic}

In real networks, traffic flowing across networks is stochastic, with the distribution of inter-arrival times being an active area of research 
\cite{garsvaPacketInterarrivalTime2014,PerformanceModelingWireless}. As suggested by the work of Garsva et al. \cite{garsvaPacketInterarrivalTime2014}, an exponential distribution is a reasonably good fit for the inter-arrival times between packets on a network. Consider the number of packets $X$ in a time period $\Delta t$ sent by a random process with exponentially distributed inter-arrival times, with a mean of $\lambda$. From basic probability theory, we know that the random variable $X$ will be Poisson distributed with a mean of $\frac{\Delta t}{\lambda}$ \cite{durrettrickProbabilityTheoryExamples2010}. Hence, when discretising our model, a Poisson distributed number of packets in a time period is a reasonable assertion. 

In our agent based model, we cannot directly assert an inter-arrival time of packets, or equivalently the number of packets in a given time period, across the network. Instead, we can control this by dictating the rate at which new packets are passed into the network by nodes. Intuitively, the traffic experienced by a node on the network is proportional to the number of paths routed through it, and the probability that a packet is sent along each of these paths. Let $s$ be the probability of a node sending a packet in one of the model's timesteps. In doing this, we suppose that the timesteps in this model represent a time period small enough that only one packet is sent per timestep. The probability of $k$ packets being sent by an endnode in a time window of $\Delta t$ is then Binomial distributed rather than Poisson. However, from basic probability theory, we know that for a large enough sample size - or equivalently a large enough number $n$ of discrete timesteps - the probability density function of the Binomial distribution approaches that of the Poisson. 

\begin{equation}
    \lim_{n \rightarrow \infty} \frac{n!}{k!(n - k)!}s^k (1 - s)^{n - k} = \frac{s^k e^{-s}}{k!}
    \label{equation:binomial}
\end{equation}

Hence, sending a packet with probability $s$ per timestep for a large enough number of timesteps is consistent with the exponential distribution of inter-arrival times we expect from a real network. To formalise the sending of traffic across the network, we define random variable $\mathcal{S}$ as representing whether a packet is sent from an endnode in some timestep $\tau$ - $\mathcal{S} = 1$ - or not - $\mathcal{S} = 0$. 

\begin{align}
    &\mathcal{S}(s) : \Omega \rightarrow \{0,1\} \\
    s.t \quad &\mathbb{P}(\mathcal{S} = 1) = s \nonumber\\
     &\mathbb{P}(\mathcal{S} = 0) = 1 - s\nonumber
\end{align}

Where $\Omega$ is the outcome space of the random variable, which is mapped to either 1 or 0. This random variable is sampled from once per endnode in each timestep to determine which nodes will send a packet in any given timestep. The target of the newly sent packet is then randomly chosen from the set of other switches.

\subsection{Representing Anomalous Behaviour}

With the introduction of stochastic background network traffic, the network model can now simulate traffic build-ups, periods of low traffic, and dynamic load balancing behaviour. In a real network, one would not have perfect information about all traffic across the system. In order to represent this, a subset of our nodes are denoted \textit{monitors}. These nodes send packets between one another along probing paths, as is common in other network tomographical studies \cite{heRobustEfficientMonitor2017,maNodeFailureLocalization2014,renRobustNetworkTomography2016,zhangNetworkTomographyApproach2018}. As these packets traverse the network, they are delayed, queued and rerouted in exactly the same manner as other packets on the network. After they arrive at their target, the delay they experience is saved, allowing for the development of delay-time histograms along each probe path as the simulation runs. One thing to note about the monitor nodes is that packets from normal switches are still sent to them when random background traffic is generated. In applications to computer networks, this simplification can be justified by supposing that nodes represent routers which connect to the switches passing on packets from individual local area networks.  

Given the aim of this work, the model also needed to permit anomalous behaviour to be inferred. Although many such behaviours may be of interest to network analysts, we settled on a simple notion of affected nodes holding onto packets for longer. In each timestep, these nodes have a chance $h$ of `holding' onto the first packet in the queue rather than passing it onto the next node, leaving a signature on the delay distributions of transiting packets.

Similar to our packet sending variable $\mathcal{S}$, we define the random variable $\mathcal{H}(h):\Omega \rightarrow \{0,1\}$ such that:

\begin{align}
    &\mathcal{H}(h) : \Omega \rightarrow \{0,1\} \\
    s.t \quad &\mathbb{P}(\mathcal{H} = 1) = h \nonumber\\
     &\mathbb{P}(\mathcal{H} = 0) = 1 - h\nonumber
\end{align}.

\begin{figure}[H]
    \centering
    \subfloat{\includegraphics[width=0.4\textwidth]{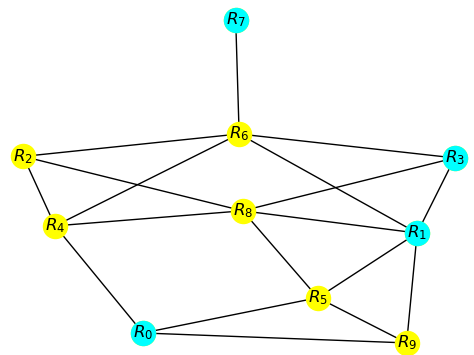}}\\
    \subfloat{\includegraphics[width=0.4\textwidth]{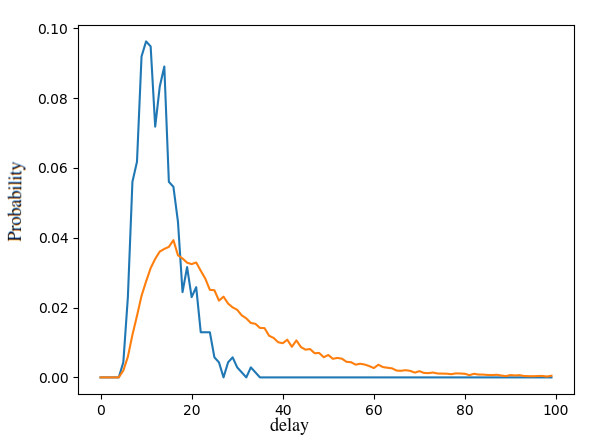}}
    \caption{Delay distributions for two different node configurations along the path $R_7 \rightarrow R_0$. The blue curve has anomalous nodes $R_0,R_1 \& R_3$, and the orange $R_7 \& R_6$. This simulation was run for $10^6$ timesteps with $s = h = 0.2$}
    \label{figure:nefexample}
\end{figure}

At each timestep, nefarious routers sample from $\mathcal{H}$. A value of $1$ means the router will propagate all packets in their queues forwards in time, but not pass them forwards on their routes. In Figure \ref{figure:nefexample}, we see that altering the number and position of nefarious routers can result in differing delay distributions between monitor nodes. 

\section{Assessing Placement Algorithms}
The purpose of this work is to compare how different monitor placement algorithms affect the feasibility of inferring the positions of problematic nodes. Three algorithms were investigated. Ma's Algorithm, as outlined in \autocite{maNodeFailureLocalization2014} is a contemporary example of a monitor placement algorithm designed for the node failure localisation problem for static networks. Given that this algorithm works by ensuring that where possible, no nodes are not traversed by paths between monitors, we expect this algorithm to continue performing well in the stochastic and load balancing case. The algorithm works by first finding fringe cases in which individual nodes could never be traversed if they were not allocated as monitors, and then playing the remaining monitors by maximising the number of nodes that are traversed by each new addition of a monitor. The second algorithm, denoted `Greedy', is the same as the Ma algorithm but without the fringe cases. It simply chooses the nodes that are farthest apart, then iteratively chooses the next monitor which causes the traversal of the largest number of new nodes. Finally, a random allocation was used as a control algorithm.

As suggested in section \ref{section:MCMC}, given an `observed' set of delay and path-length distributions, the solution space of anomalous node locations, hereafter referred to as candidates, can be explored. A direct way of comparing algorithms would then be to see how well MCMC performs in each case. However, the large variance in MCMC convergence time and steps spent on the true solution means that a large number MCMC runs would need to be performed for statistical significance. A more robust, and less computationally intensive method based on the solution space was then developed.

\begin{figure*}[p!]
    \centering
    \subfloat[Ma]{
        \includegraphics[width=0.45\textwidth]{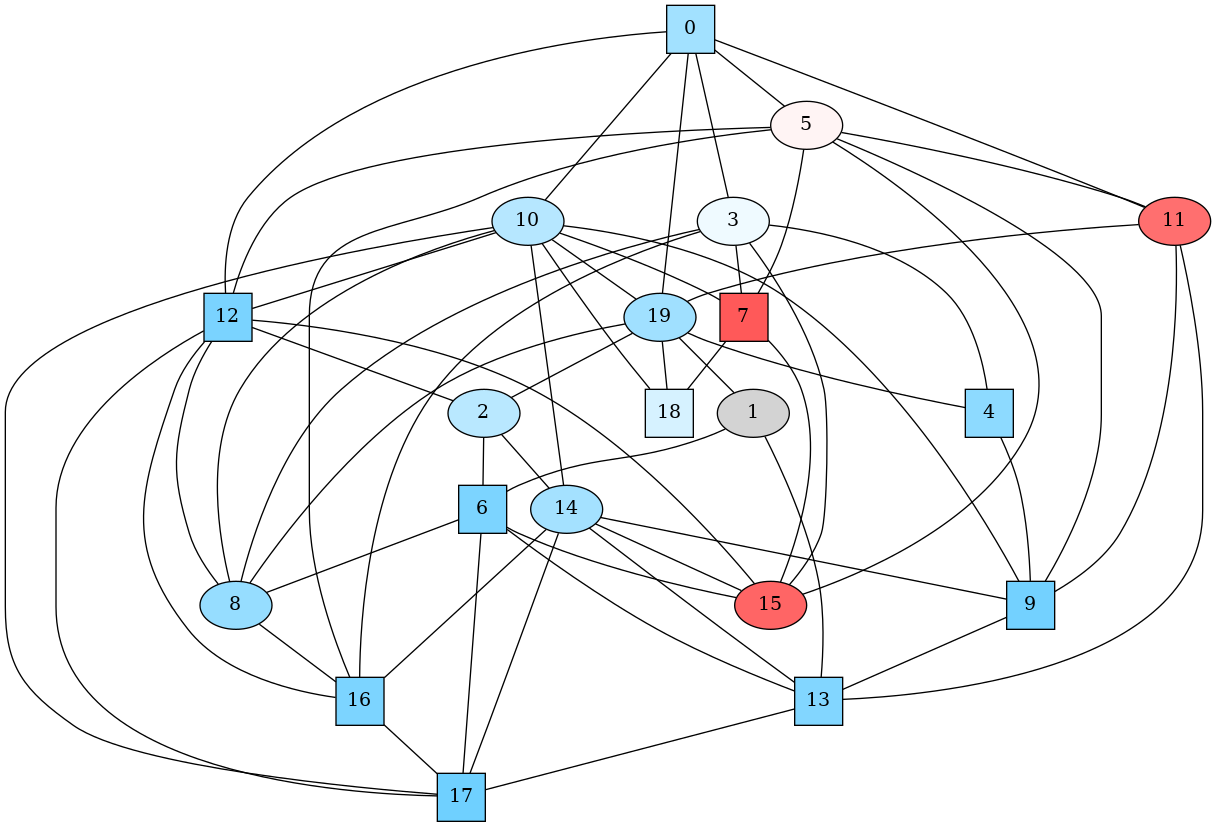}
        \label{figure:heatmap:ma}
    }
    \subfloat[Greedy]{
        \includegraphics[width=0.45\textwidth]{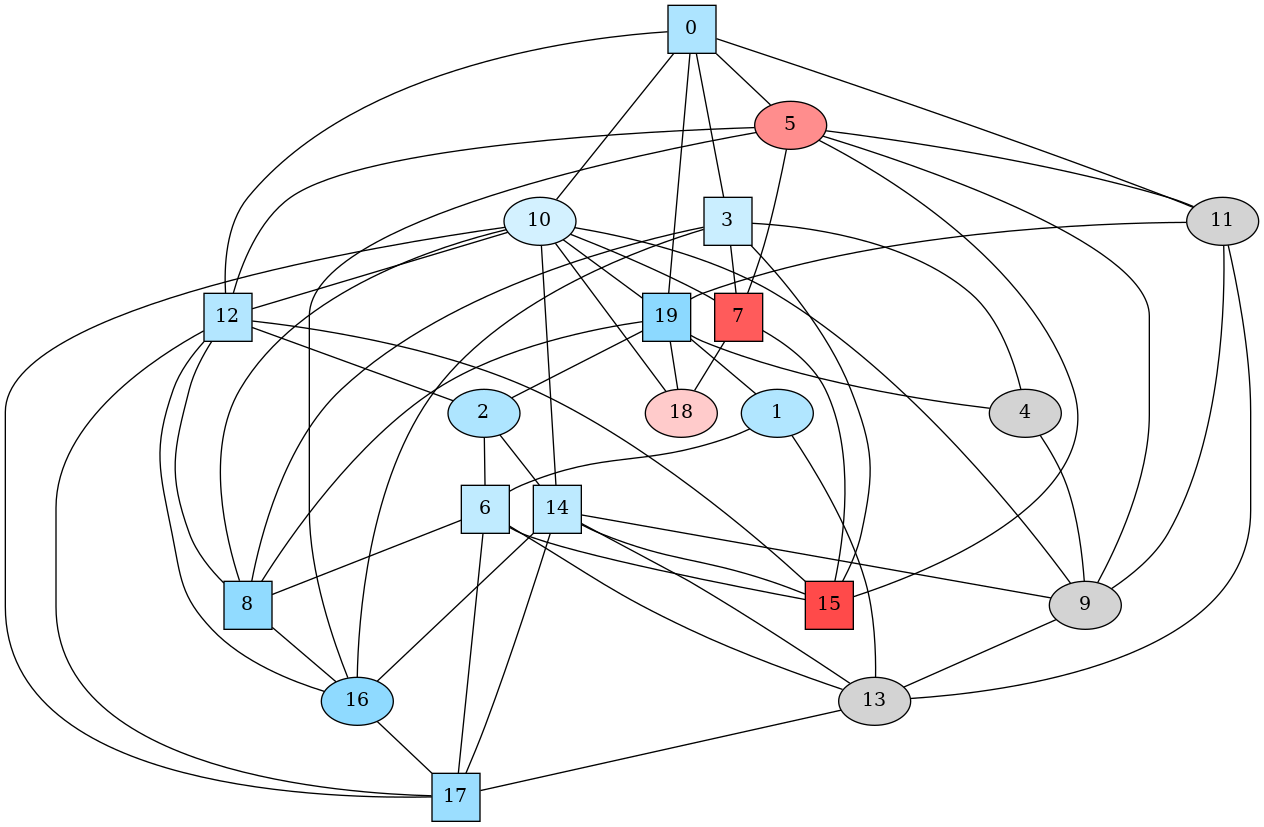}
    } \\
    \subfloat[Random]{
        \includegraphics[width=0.45\textwidth]{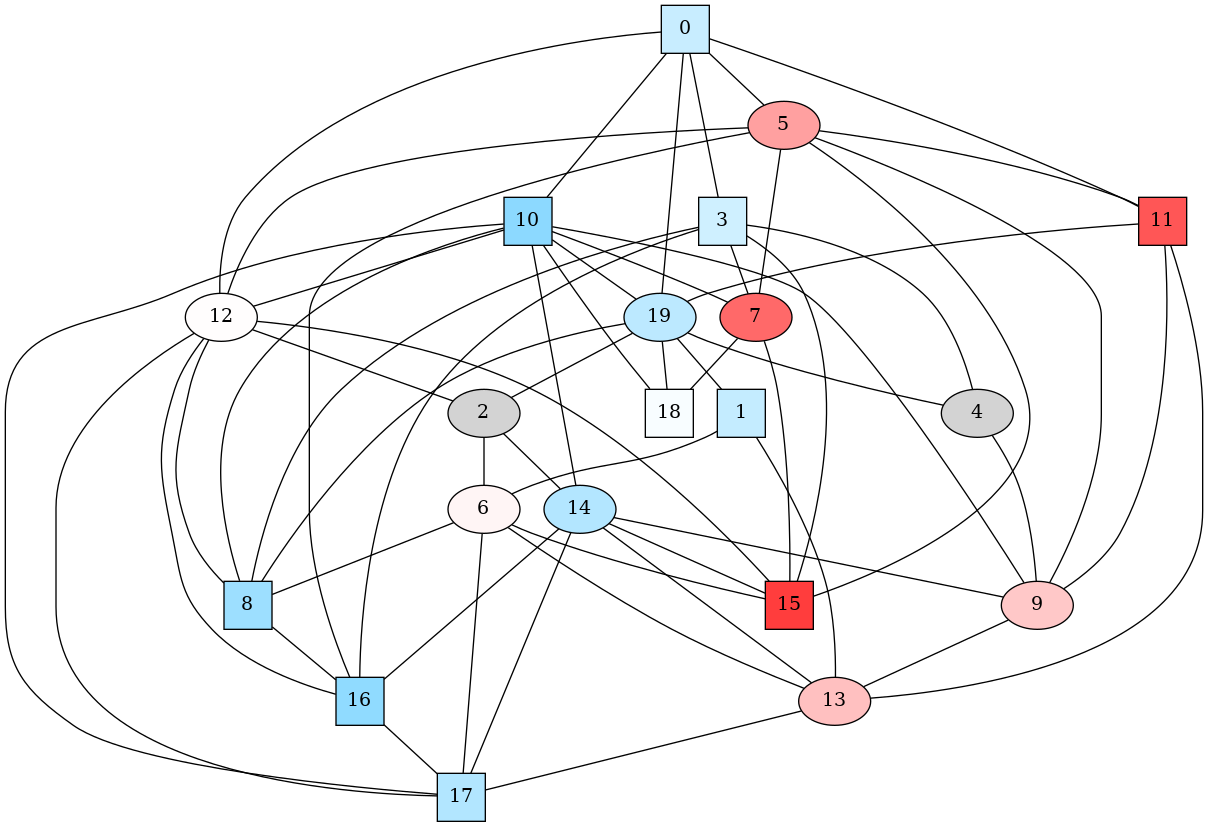}
    } 

    \caption{The `heatmaps' for the different placement algorithms showing the effect of making each node nefarious on the solution space, averaged along each path. Red and blue correspond to better or worse matches to the observed distribution respectively, with the colour map normalised about the largest result. The true configuration is $\{7,11,15\}$ which is shown most prominently by Ma's algorithm. Figure \ref{figure:heatmap:grid} shows the effect on each path between monitors for the Ma example.}
    \label{figure:heatmap}
\end{figure*}

\begin{figure*}[p!]
    \centering
    \includegraphics[width=0.6\textwidth]{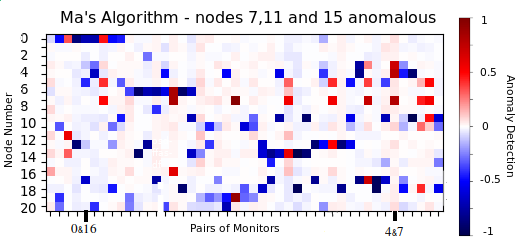}
    \caption{The underlying results used for creating the heatmap graph for Ma's algorithm in Figure \ref{figure:heatmap:ma}. The colour of row $i$ and column $j$ represents the effect of making node $i$ anomalous on the distributions across path $j$. For the sake of readability, all but two example paths on the x axis have been omitted.}
    \label{figure:heatmap:grid}
\end{figure*}

For MCMC to perform effectively, we require that the solution space has a well defined minimum about the true configuration. Otherwise the algorithm will struggle to distinguish the correct set of anomalous nodes from false sets, let alone converge to it. A more robust way of comparing the effectiveness of placement algorithms is then to quantify the suitability of the solution space. To demonstrate this concept, we ran the model on a small sample network with randomly chosen sets of anomalous nodes, generating an observed set of distributions $\mathcal{D}_{Ob}$. The model was then run with all healthy nodes, giving a set of `control' distributions between monitor pairs $\mathcal{D_H}$. Then, each node was` chosen one at a time, generating distribution sets $\mathcal{D}_n$, where the subscript $n$ corresponds to the $n$th node being anomalous. For each node, the difference between $\nu(\mathcal{D}_{Ob} , \mathcal{D}_n )$ was computed, and subtracted from $\nu(\mathcal{D}_{Ob} , \mathcal{D_H} )$ to assess changes to the solution space between the `healthy network' scenario and single anomalous routers. These `heatmaps' are shown in Figure \ref{figure:heatmap}. The colour scale indicates the effect of making a particular node anomalous when compared to the control or `healthy' network. White suggests that  there is  little discernible difference, while red and blue suggest that the delay distributions become more, or less, similar to the observed, respectively. The colours on the heatmap networks correspond to the average comparison of distributions along every pair of monitors, with an example of this underlying process shown in Figure \ref{figure:heatmap:grid}. In this grid of the heatmap, we see that the path between monitor $0$ and $16$ correctly detects that node $5$ is healthy, while that between $4$ and $7$ picks up $3$, $4$ and $7$ as anomalous. It's worth noting that $3$ and $4$ aren't in fact anomalous, but since $7$ is, the addition of a single anomalous node between $7$ and another monitor will appear to improve the distribution. The false positives are balanced out by other paths between monitors, which detect that $3$ and $4$ aren't anomalous.  By comparing the Figures in \ref{figure:heatmap}, it is clear that the choice of monitor placement algorithm affects the solution space of candidate sets of anomalous nodes.

\end{multicols}

For this network configuration and choice of nefarious routers, it is clear that correct anomalous nodes - $\{7,11,15\}$ - show an improvement over the `healthy network' case, whereas almost every other node causes the distributions to stray further from the observed. In comparison, the Greedy algorithm could detect no change for node 11, but found a weak positive effect for node 5, and the random placement resulted in a yet weaker result. However, this sample represents just one network and one monitor configuration, and other samples show weaker results for Ma's algorithm. For a more comprehensive analysis, we need to aggregate over a much larger sample. 

To do this, we first randomly generated a 20 node network, and chose up to 3 random nodes to be anomalous. A placement algorithm was then used to place monitors, and the model run to generate our observed distributions $\mathcal{D}_O$, and true distribution $\mathcal{D}_T$. Note that the only difference between these two distributions are due to the stochastic behaviour within the model, meaning that we expect $\nu( \mathcal{D}_O,\mathcal{D}_T)$ to be small but non-zero. We then define the suitability of the solution space $R$ as the mean difference between $\nu( \mathcal{D}_O,\mathcal{D}_C)$ for each $N$ incorrect candidate configurations $C \in [1,N]$, and $\nu( \mathcal{D}_O,\mathcal{D}_T)$.

\begin{equation}
    R = \frac{1}{N}\sum^N_{C = 1} \frac{\nu( \mathcal{D}_O,\mathcal{D}_C)}{ \nu( \mathcal{D}_O,\mathcal{D}_C)}
\end{equation}

\begin{figure}[H]
    \subfloat[MCMC Performance]{
        \includegraphics[width=0.9\linewidth]{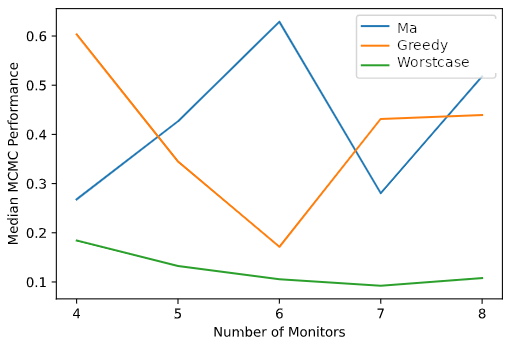}
    } \\
    \subfloat[Proxy Performance]{
        \includegraphics[width=0.9\linewidth]{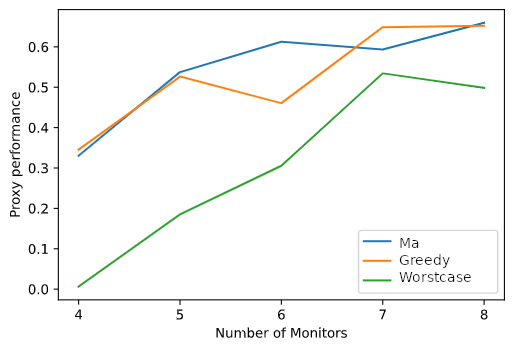}
    }
    \caption{Verification that proxy technique corresponds qualitatively to MCMC performance}
    \label{figure:compareproxy}
\end{figure}

However, the validity of assessing MCMC's performance with this proxy technique still needed verifying. To assess whether a higher value for $R$ corresponded to a faster MCMC convergence time, a dozen different anomalous router configurations were randomly chosen and 50 MCMC runs of 10,000 steps were performed for each. The proportion of the MCMC steps spent at the correct configuration of nodes, normalised to the highest value, was calculated, and the median plotted against the corresponding proxy performance. The sample size was necessarily smaller due to the large computational task of executing the model in conjunction with a MCMC algorithm. This meant that the random monitor placement used as a control would be a poor way of providing a reference point for MCMC performance, as there wouldn't be enough samples for the random behaviour to be averaged over. Instead, a `worst case' algorithm was used, which places monitors deliberately next to one another so as to cover the smallest amount of the network as possible. It is then expected that this placement algorithm should perform worse than the nuanced ones. The results as shown in  \ref{figure:compareproxy} demonstrate that performance of MCMC is qualitatively consistent with the proxy. It is noteworthy that the proxy does not distinguish the two nuanced monitor placements from one another as much as MCMC performance. Differences are to be expected between the two methods for comparing monitor placements. One reason for this is that the proxy approach averages over many areas of the solution space to determine how conducive it is to MCMC convergence, but once convergence occurs, the solution space local to the solution becomes more relevant. 

\begin{figure}[H]
    \includegraphics[width=0.9\linewidth]{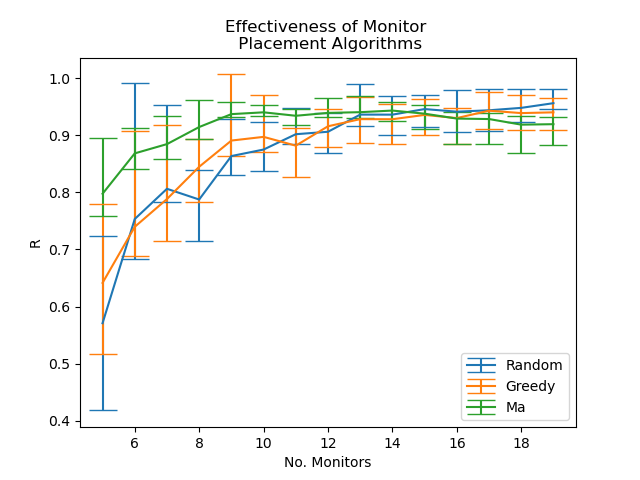}
    \caption{Aggregated performance of the monitor placement algorithms}
    \label{figure:placements}
\end{figure}

To compare monitor placements over a large sample size, the proxy method was used to compare the performance of the monitor placement algorithms for 400 candidates and 24 random 20 node networks for various numbers of monitor nodes. As demonstrated by Figure \ref{figure:placements}, Ma's algorithm consistently performs better than the others when there are few (4 to 10) monitors. As expected, once the number of monitors increases towards the number of nodes in the network, the choice of monitor placement has a smaller effect once a significant proportion of nodes are monitors. 

This finding suggests monitor placement theory for fixed networks can be extended to dynamic, load balancing networks, with Ma's algorithm giving the best result. In addition, a novel Markov Chain Monte Carlo approaches has been demonstrated to work as a dynamically routed network inference tool. 

\section*{Conclusion and Further Work}

This work demonstrates that Markov Chain Monte Carlo techniques can be used for the stochastic Node Failure Localisation problem for load balancing networks. Further, the results suggests that more nuanced monitor placement algorithms, specifically Ma's algorithm, result in better inference performance. A natural extension to this work would be to assess the effect of network design in addition to monitor placement on the performance of inference. A variety of `anomalous' behaviours could also be explored in addition to traversal time delay to investigate the scope of MCMC methods in stochastic networks.

\printbibliography


\end{document}